# Data Trust and IoT

Mayra Samaniego
University of Saskatchewan
mayra.samaniego@gmail.com

## ABSTRACT

People's IoT surroundings have become a valuable source of information that can impact individuals and society positively. An individual's IoT data can be used for different purposes – for instance, research and improvement of public services. However, individuals lack governance power to share their IoT data. Data trust is a concept that brings opportunities in addressing data sharing in IoT. This research reviews the concept of data trust. Then, we review IoT and its unique characteristics that make the implementation of data trust a challenge. We further discuss blockchain technology and how it can be used to enable data trust in IoT. Finally, we introduce a blockchain-based solution for data trust in IoT.

## 1 Introduction

In the online environment, where a vast amount of information is shared and transferred around the world in near real-time, it becomes increasingly difficult for individuals to maintain control over how their data is shared, used, and the benefits that can be achieved from it.

There is an increasing demand from individuals to prevent others from accessing their IoT data without their consent. Also, there is an emerging demand for mechanisms that give individuals decision-making power over the data their IoT systems generate. The following are some of the things that individuals are most concerned about when sharing their IoT data:

- Transparency
- Digital and physical privacy
- Traceability

Additionally, the data that IoT systems generate has the distinctive characteristic of being transmitted in near real-time (data streaming). The following are some considerations when sharing data streaming:

- Data constantly flows through a series of components
- A single piece of data is worthless
- Timely, preventive, and reactive decision making

Not much has been studied about data trust for data sharing. Data trust is a data sharing concept that seeks to appoint a steward (aka trustee) to manage data for a purpose on behalf of a beneficiary or beneficiaries that own the data [1]. Data trust establishes the rights and obligations that trustees and data owners get. The concept of data trust can be used to enable a mechanism that allows data owners to make free decisions on how they share their data. A data trust platform would provide the following benefits when sharing data:

- Encourage interaction between people anywhere in the world
- Promote digital dialogue between data owners and third parties that are interested in using that data.
- Provide a context in which to add rules and comments and provide clarity on what this information really means to different people.
- Provide an environment in which it is possible to audit how the data has been shared and used.

When designing and implementing data trust solutions in IoT some challenges emerge. This research reviews the characteristics of IoT, and the data sharing demands. Also, this research proposes a data trust platform using blockchain.

## 2 IoT

The internet of things (IoT) is a network of common devices and objects connected among them and to the internet to enable communication and interaction between the physical world and people [2]. IoT collects data, processes it, and makes use of it to detect something and do something intelligent [3]. The term internet of things emerged for the first time in 1999. Kevin Ashton coined it in a presentation about linking the internet to supply chain management [4]. Nowadays, the advent of IoT has led to the development of solutions for different fields, such as healthcare [13, 14], smart spaces [15, 16], public services [17, 18], industry [19, 20], among others.

IoT is an active area of research. Contemporary researches have pointed out challenges about data sharing and how the benefits of IoT data and its derived data are managed [13][14][15][16]:

- What happens to the IoT data collected?
- Who controls the benefits of the collected data?
- Who controls the benefits of the derived data?

## 2.1 Challenges of IoT

IoT is cooperation among different hardware and software technologies [17]. The main goal of IoT is to connect what has not been connected before to enable an intelligent network. Thus, more control of the physical world is enabled, and more advanced applications are developed to enable rich communication and interaction between objects, systems, and people. The principal difference among traditional internet and IoT is the generation of data and how data is used [18]. Some of the characteristics that define IoT are listed below:

- Pervasive Networks: IoT is highly pervasive. IoT enables connectivity at the constrained level, locally. IoT can get the data from anywhere, at any time, and under different environmental conditions, and humans do not even notice it. For example, wearable devices.
- Heterogeneous Environment: This heterogeneous environment seeks to enable communication with a wide variety of devices using multiple communication technologies – for instance, RFID, Bluetooth Low Energy (BLE), Zigbee, or 5G.
- Heterogeneous Application Communication Protocols: IoT enables communication by implementing multiple application communication protocols – for instance, DDS, MQTT, and CoAP (these protocols will be explained deeply in section 1.7).
- Heterogeneous Data Streams: IoT produces heterogeneous data streams. The variety of IoT data generated involves structured, unstructured, and semi-structured data. Also, the variety of data might involve different interpretations depending on the context of the IoT scenario [19]. The velocity of the generation of data allows us to execute near real-time analytics and make decisions.
- Low Data Rates: IoT constrained networks support low data rates and tend to be lossy.
- Dynamic Interoperability: Device identification and interoperability when having heterogeneous devices.
- Changing Network: IoT is continuously changing. IoT devices are plugged in and removed regularly. The state of IoT devices can change dynamically. e.g., sleeping and waking up, connected, and/or disconnected — also, the context of devices changes, including location and speed.
- Interaction Patterns: IoT interaction patterns continuously change. A lot of these patterns are interactions with externally hosted services.
- High Scalability: IoT demands high scalability. IoT demands support for the increasing number of connected devices, data, analytics, applications to adapt to changes in the environment and meet the expectation of users without degradation in the quality of the service.

This research focuses on the challenges that emerge when sharing IoT data in the context of a smart home. In a smart home context, the following facts are known and obvious:

- When data is being collected
- By which IoT devices
- By which organization
- For what purposes.

However, when device owners exercise their rights to consent to data sharing, the following concerns emerge [20][21]:

- What further benefits are been obtained from the collected IoT data?
- How access to data can be enabled integrating specific variables, for instance, operations allowed
- How data is being used by others
- Users want to understand what data is being derived from their original data
- Users want to receive the benefits of derived data

IoT has to guarantee trust when data is collected, maintained, and shared to realize its full benefits.

Data trust is a data sharing methodology that can help address the previous questions and data sharing challenges

## 3 Data Trust

Data trust represents a way to share data in a trustworthy manner. According to Rieder et al. [22] trust is the key to build and maintain a social order as without trust societies might collapse. However, what we trust and the circumstances in which we trust change over time – for instance, nowadays, people trust the numbers and statistics obtained by sensors and big data.

A preliminary work presented by the Oxford Internet Institute [23] suggested that the identity of entities, privacy, and security represent main aspects of online trust. Trust issues emerge from the difficulty to identify the identity of users and online services and business. Additionally, Hoffman et al. [24] suggest that privacy is the main concern when developing online trust.

According to Edwards L.[1], trust is managed by an entity that allows a trustor to transfer property to a trustee. The trustee gets rights thereafter to use the data, but the trustee gets restrictions ruled by the terms of the trust and the rights and best interests of the beneficiaries (figure 1).

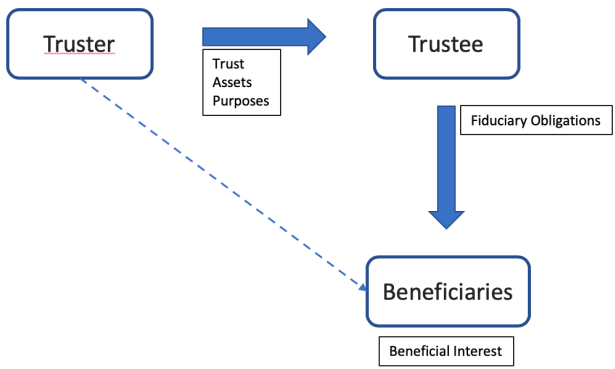

**Figure 1. Overview of a data trust relationship** [1]**.**

The data trust concept and how it can be implemented to enabling data sharing have been poorly studied. Previous studies have focused on the study of trust to enhance consumers' online trust [25][26], management of event detection in networks [27], management of public services [28]

Based on the definition of trust presented by Edwards L.[1], when a data trust relationship is established, there are three identified actors:

- Trustors
- Trustees
- Beneficiaries

There are three key features of a data trust relationship:

- Transparency
- Value delivery
- Consequence acceptance

Transparency is the main feature that enables trustworthiness. Transparency when sharing, looking after, and making decisions about data.

Value delivery refers to distribute the benefits that arise from data equitably and giving the maximum value to the actors on a data trust relationship. The value delivery is mostly monetary.

Consequence acceptance refers to clarifies the repercussions that misuse of data can cause. Trustors and trustees must agree exactly about what is allowed to do over data. Also, trustees must do exactly what was agreed.

Based on these previous implementations of data trust, this research states that to enable data trust, the following requirements have to be addressed:

- A mechanism to identify agreement and define shared objectives among data trustors and trustees.
- A mechanism to allow data owners and organizations to define the purpose for which the data will be used.
- A mechanism to define the specific rights, obligations, and limitations that organizations have on the shared data.
- A mechanism to allow data owners and organizations define the benefits when the data is shared and used to create a value.
- A mechanism to allow data owners and organizations to formally create and validate a data trust relationship that includes all the previous explained conditions.
- A mechanism to allow data owners and organizations validate compliance with the conditions of the data trust relationship.
- A mechanism to allow data owners and organizations to end their data trust relationships.

A data trust relationship should include the following characteristics:
- A clear purpose
- A legal structure that includes the previously explained actors
- The rights and duties over data
- The benefits and how they are shared
- The consequences for misuse of data

Any third party willing to use and individual's data must state exactly what it will do. Once, these actions are agreed the third party must do exactly what was agreed.

## 3.1  Data Trust and IoT

IoT has to guarantee trust when data is collected, maintained, and shared to realize its full benefits.

In the context of smart homes, data trust is seen as a potential way of giving users more control when it comes to the sharing of their collected data.

This research explores the idea of using blockchain technology to create a data trust solution to enable data sharing in IoT. This research proposes that IoT data can be shared by a data trust implemented in the form of a blockchain technology.

Data trust for data collection and data sharing.

Khan et al. [29] presents the requirements that should be addressed to gain consumer trust when creating IoT systems. This study states that privacy requirements emerge at different stages of interactions and integrate four questions of privacy: what, where, who, and when. Consumers and vendors must what exactly IoT devices sense, where that sensed data is stored, when the data is being used, and who can access to the data.

Access control relies on policies to determine how data is shared [30]. Thus, it limits what users can do directly or indirectly to prevent security issues when sharing their data. Traditional internet management develops policies centrally. A central administration entity is the one that owns the network and manages the level of access that users have.

Table 1 presents an overview of IoT data interactions.

**Table 1. Data Interactions in IoT**

| Parameter | Traditional Internet | IoT |
| --- | --- | --- |
| Data creation | By humans | By sensors |
| Data consumption | By request (search engines) | By pushing information and triggering actions |
| Data flow | Through links that connect web resources | Through defined operators that connect devices and constrained services |
| Data Value | Answer users' questions | Action and timely information. The value is added by analytics. |

## 4 Blockchain and Data Trust

Blockchain is a distributed ledger that enables the execution of direct transactions without any central verification authority and stores transactions in the form of blocks [31]. The main characteristics of blockchain technologies are:

- Immutability. Blockchain networks are tamper-proof. Previously written blocks cannot be changed, which makes them immutable.
- Reliability. Blockchain technologies execute a validation process to validate transactions before writing them on the chain. This validation process is called consensus.
- Provenance. All transactions are recorded in-chain.

This research explores the use of blockchain technology to enable data trust in IoT. Blockchain technology can provide the high level of transparency, value delivery, consequence acceptance that data trust demands. Thus, data owners can control who they share their data with.

The following features of blockchain guarantee transparency:

- Blockchain distributes all transactions towards the ledger. The core of transparency is represented by the large blockchain network having identical copies of all the transactions.
- A blockchain ledger is immutable, which means that it remains unaltered and indelible.
- In a blockchain network all transactions are visible. Anyone can query previous executed transactions.

The following features of blockchain guarantee value delivery:

- The distributed ledger of a blockchain technology is formed by nodes located in different places around the world. Instead of single entities establishing a data trust relationship, it is established by the distributed network.
- Blockchain does not have any single point of failure, making it more resilient, efficient and democratic.
- Transactions are grouped in a block, which is secured by a cryptographic hashing process that generates a hash pointer.
- This hash pointer contains the hash of the current block and the hash of the previous block. This approach ensures that all the blocks are linked retroactively. If somebody tries to modify the data contained in a block, this modification will generate a totally different hash pointer. Thus, in order to modify the transaction, you would have to redo the block's proof of work, i.e. remine the block. Modifying transactions already in the blockchain requires remining blocks, and after a transaction already has a few confirmations, doing this requires immense amounts of computing power. So much computing power is required that modifying blocks is effectively impossible to do.
- Transactions are validated through a consensus mechanism that is executed by all the blockchain network.

The following features of blockchain guarantee consequence acceptance:

- A blockchain ledger is immutable, which means that it remains unaltered and indelible.

Additionally, blockchain associates a pair of keys (aka addresses) to process each transaction in the block. Blockchain associates the public key with the transaction and keeps it visible to all. On the other hand, blockchain requires the private key to authorize the execution of the transaction. The owner of the blockchain account is the only one that should know the private key. On the ledger that blockchain provides, any transaction can be queried by the public key. However, connecting the public to the account owner is nearly impossible as the private key is encrypted.

Contemporary researchers have analyzed how blockchain would benefit IoT [191, 192] The following paragraphs review how some researchers have integrated blockchain technologies to create systems to manage different aspects of IoT.

Rifat et al. [34] follow the ADEPT idea of creating hybrid architectures that integrate existing technologies. This work proposes a blockchain-based IoT infrastructure that integrates LoRas for communication with IoT devices, Swarm for distributed data storage, and Ethereum to store the backend services in the form of smart contracts. Data about IoT devices is stored in Swarm. This approach allows using the public Ethereum network instead of configuring a private one. Ethereum stores the smart contracts that manage access rules and permissions.

Samaniego et al. [35] also present a hybrid system for autonomous IoT management. This work integrates CLIPS to enhance the constrained network with artificial intelligence features and Multichain to handle communication. This work introduces the internet of "smart things" – IoST. Smart things are software artifacts provisioned with artificial intelligence features. These features are programmed in the form of rules using the CLIPS programming language. The smart things can autonomously analyze their state, infer knowledge, and monitor changes.

Sheng et al. [36] propose a blockchain-based system that follows the attribute-based access control. IoT devices are defined as a set of attributes. This work implements Hyperledger Fabric to record the attribute-based permissions. Attribute authorities record the authorization of the attributes of the IoT devices and the permissions associated with them in the form of transactions.

Novo O. [37] proposes an architecture that integrates blockchain to handle roles and permissions to access IoT devices. This architecture implements a private Ethereum network that manages access through a single, smart contract.

Samaniego et al. [38] implement a blockchain-based system that supports multi-tenant access in IoT. This research presents the design of virtual resources that are hosted among edge devices. These virtual resources enable the definition of views on top of existing IoT systems. Each tenant gets its own virtual IoT system. This approach offers virtualization at the edge of the constrained networks without the high latency that traditional cloud-based systems that support multitenancy involve. This work implements IBM Bluemix to store the virtual IoT systems for each tenant.

Rifi et al. [39] propose a blockchain-based architecture to Uses Ethereum smart contracts for authenticating users just once. After users have been authenticated, the smart contract broadcasts an access token among the IoT network. IoT devices can manage access based on the copy of the token they have. Thus, users do not have to go through the smart contract every time they want to access the IoT network.

Ouaddah et al. [40] present an access control framework for authorization management. This work integrates Ethereum smart contracts to evaluate contextual information and apply policies to make authorization decisions.

Pinno et al. [41] present a blockchain-based architecture to store relationships between users or groups of users and devices to enable access to IoT data. This work defines two types of relationship references, blockchain-dependent and external. The blockchain-dependent relationship links information about users with information about their authorization stored in a block. This approach helps to audit provenance information. The external relationship links information about users with information about their last access to the system. This approach helps to improve latency when confirming the permissions of users.

## 4.1 Smart Contracts

Smart contracts represent the second generation of blockchain technologies [42]. Ethereum is the main blockchain technology that represents this generation. Ethereum was proposed by Vitalik Buterin in 2013 [43]. It was formally launched in 2015. Ethereum is well-known for implementing smart contracts. Ethereum is a public and open source blockchain technology that allows deploying smart contracts to build distributed systems without a third-party orchestration. This approach replaces powerful centralized servers by a distributed large network of small computers located around the world running smart contracts.

Nick Szabo initially proposed smart contracts in 1994 [44]. Szabo states that smart contracts are computerized protocols that execute the conditions of a contract. Specifically, he defined a smart contract as: *"a set of promises, specified in digital form, including protocols within which the parties perform on these promises."*

Ethereum defines smart contracts as "systems that automatically move digital assets according to arbitrary pre-specified rules" [43]. Smart contracts directly manage the transfer of assets between parties after these parties have agreed to specific conditions. As long as the conditions are met, the result of a smart contract cannot be stopped. For instance, a smart contract code automatically validates conditions and determines what to do with an asset. The smart contract determines whether the asset should be transferred to a new person or whether it should be returned to the person who sent it originally.

Smart contracts in blockchain facilitate the agreement between untrusted parties without relying on am unique centralized trusted third party. Smart contracts control the execution of transactions, and blockchain ensures that those transactions are trackable and irreversible.

Ethereum smart contracts are deployed on the Ethereum virtual machine. The computing power to run smart contracts is provided by all the nodes that are part of the network. Once a smart contract is deployed, no one can alter it. A smart contract may be organically stopped if it has an ending function and if the condition of the ending function is met. Smart contracts can facilitate the creation of data trust relationships. For instance, to create an online data-trust service, we can set up an Ethereum smart contract that would have the specifications of the actors and the conditions of the relationship. When the smart contract is signed by the actors and initial conditions are met, then access to data is enabled. If an actor violates some of the conditions, then some actions are triggered.

In order to deploy smart contracts in Ethereum, we need Ethers. Ether is the cryptocurrency that Ethereum implements. While smart contracts help reduce bureaucracy of processes, depending on the complexity of the smart contract, the costs of deployment and transactions might be high.

An alternative to smart contracts is chain code that Hyperledger Fabric implements. Unlike Ethereum smart contracts that run on the distributed Ethereum network, Hyperledger chain code run under the control of a peer. A specific peer managed the chain code. Unlike Ethereum smart contracts that are immutable, Hyperledger chain code are flexible because it is managed by a single peer that can update its state. While Ethereum is more oriented for mass consumption, Hyperledger is oriented for flexibility in business.

Table 2 presents comparison of the main features of Ethereum and Hyperledger blockchain technologies. Table 3 presents a comparison among private, public, and hybrid blockchain networks.

**Table 2. Ethereum vs Hyperledger**

| Attribute | Ethereum | Hyperledger |
|---|---|---|
| Participation and confidentiality | Public Permissionless | Private Permissioned |
| Consensus | Proof of Work (PoW) | Execute-Order-Validate Zab Raft |
| Cryptocurrency | Ether | N/A |
| Accounts and Identity | Execute transactions (senders and recipients) Externally owned Contract | Membership Service Provider (MSP) No pre-defined recipient |

**Table 3. Public, private, and consortium blockchain networks for data trust.**

| Attribute | Ethereum Smart Contracts | Fabric Chaincode |
|---|---|---|
| Transaction Privacy | No | Partial |
| Value Delivery | Yes | Partial |
| Consequence Acceptance | Yes | No |
| Flexibility | No | Yes |
| Deployment Environment | Distributed (EVM) | Centralized (A peer) |
| Programming Languages | Specialized (e.g.: Solidity) | High level (e.g.: Java, JavaScript, Go) |

## 4.2　Blockchain Networks

Buterin V. [45] classifies blockchain technologies in three categories, public, private, and consortium.

*4.2.1. Public Blockchain*

A public blockchain network is decentralized. Therefore, no single entity controls the network. In public blockchain networks, the level of access is granted to all nodes in the network. Any node can read and write transactions. Also, any node can participate in the consensus process. Public blockchain networks provide low scalability as all the nodes have to validate transactions, and transactions are processed at a slow rate. Public blockchain networks are trustless. Participants must not be trusted. Public blockchain networks are more secure than private blockchain networks. The decentralization and constant participation of nodes would make the network secure. Public blockchain networks would consume a high amount of energy as all nodes compete against each other to validate blocks.

*4.2.2. Private Blockchain*

A private blockchain network is centralized. There is a single entity that controls the network. In a private blockchain, the level of access is granted to certain nodes in the network previously authorized by the central entity. Only these pre-authorized nodes can participate in validating and writing transactions. Also, approved participants can read from the chain. Also, only these pre-authorized nodes can participate in the consensus process. Private blockchain networks provide high scalability as a few nodes handle the validation process, and transactions can be processed at a higher rate than in public blockchain networks. Private blockchain networks are not trustless. Pre-authorized nodes rely on the credibility of each other. A private blockchain network might offer less reliability than public ones. The close environment of validator nodes would make it easier to expose them to hacks, risks, and data manipulation that might compromise the entire network.

*4.2.3. Consortium Blockchain*

Consortium blockchain networks combine features from public and private blockchains. The main difference is that in consortium blockchain networks, the level of access is granted by a group of entities. Pre-authorized nodes are all known and belong to all these entities. Table 4 and 5 present an overview of these groups of blockchain [46].

**Table 4. Technological features of public, private, and consortium blockchain networks.**

| Attribute | Public Blockchain | Private Blockchain | Consortium Blockchain |
|---|---|---|---|
| Permission-less | Yes | Yes | Yes |
| Permissioned | No | Yes | Yes |
| Immutable | Yes | Yes | Yes |
| Centralized | No | Yes | Partial |
| Consensus | Proof-of work | Byzantine fault tolerance (BFT) Proof-of-stake | Byzantine fault tolerance (BFT) Proof-of-stake |
| Consensus Participants | All miners | Pre-determined group of nodes | Pre-determined group of nodes |
| Governance | Open to any entity on the internet | Single entity | Group of entities |
| Read Permission | Public | Public or restricted | Public or restricted |
| Efficiency (transactions troughput) | Low | High | High |
| Performance | Higher costs Slower transaction speeds compare to private blockchains | Cut down costs Increase transaction speeds | Cut down costs Increase transaction speeds |
| Example Technology | Bitcoin Ethereum | Ethereum | Ethereum Hyperledger Fabric |

**Table 5. Public, private, and consortium blockchain networks for data trust.**

| Attribute | Public Blockchain | Private Blockchain | Consortium Blockchain |
|---|---|---|---|
| Transparency | Yes | No | Partial |
| Value Delivery | No | Yes | Partial |
| Consequence acceptance | Yes | Yes | Yes |
| Anonymity | Yes | No | Partial |
| Transaction Privacy | No | Yes | Partial |

# 5 Proposed Data Trust Platform

We can use blockchain beyond cryptocurrencies. The design pattern and the characteristics of blockchain technology can enable a decentralized management structure to manage IoT networks[47]:

- A decentralized system to establish rules.
- A copy of every transaction executed to guarantee transparency and auditing.
- A decentralized validation of transactions to guarantee the reliability of the business.
- A tamper-proof ledger to guarantee the security.

The ability to create, store, and transfer digital assets in a distributed, decentralized, and tamper-proof manner is of a large practical value for IoT.

The ADEPT (Autonomous Decentralized Peer-to-Peer Telemetry) system [48] was the first system to integrate blockchain technology to manage IoT networks. The ADEPT system was developed by IBM and Samsung. ADEPT provides a hybrid architecture that eliminates the classical centralized management authority in IoT. The main goal of ADEPT is to build a decentralized IoT management structure by providing a hybrid architecture. ADEPT supports the following management functions:

- Peer-to-peer messaging by integrating Telehash. Telehash is an open-source distributed hash table implementation (DHT) of the Kademlia protocol. Telehash is used to manage notifications among devices.
- Distributed file sharing by integrating BitTorrent. BitTorrent is a DHT file sharing. BitTorrent is used to distribute content among devices.
- Autonomous device coordination by integrating Ethereum.

ADEPT integrates Ethereum to build a device-coordination framework to handle the following transactions among devices:

- Checklists. Thus, devices can maintain their status to prevent failures.
- Contracts. Thus, devices can coordinate agreements about actions, controls, and complex transactions that require the exchanging of resources to receive a service. Contracts mainly manage the following transactions:
  - agreements
  - payments
  - barter
- Rules of engagement. Thus, devices can handle complex interactions that require the definition of rules. These rules can be based on:
  - proximity (e.g., physical, social, or temporal)
  - consensus (e.g., selection, validation, or blacklisting)
  - triggered actions by other devices simultaneously
- Authentication. Thus, devices can handle individual coordination functions like registration and authentication
- Registration. Thus, devices can autonomously handle access verification before registration of updates installation.

Ethereum provides an environment with the following characteristics:

- Trust-less network
- Signed transactions
- Public consensus
- Transaction code

The following characteristics define a smart contract:

- Self-verifying
- Self-enforcing
- Self-executing
- Tamper-proof

Smart contracts can handle the following tasks:

- Automate manual processes.
- Ensure transparency.
- Eliminate relations to central trusted parties.
- Support multi-signature conditions in order to execute a transaction.
- Manage agreements among unknown parties.
- Provide input to other contracts.

Following the ADEPT idea, this research proposes the development of a hybrid decentralized data trust platform integrating Ethereum blockchain. Figure 4 presents an overview of the architecture of the data trust platform that this research proposes. The architecture flow of the data trust platform starts with two parties interested in enabling a data-trust relationship. The two parties define the parameters that will rule the data trust relationship. The data trust platform has five components:

- Ethereum Blockchain Network.
- Off-Chain Distributed Data Storage.
- Data Trust Factory.
- Data Trust Reviewer
- Data Provisioner
- Data Streaming Factory
- Monitoring Tool

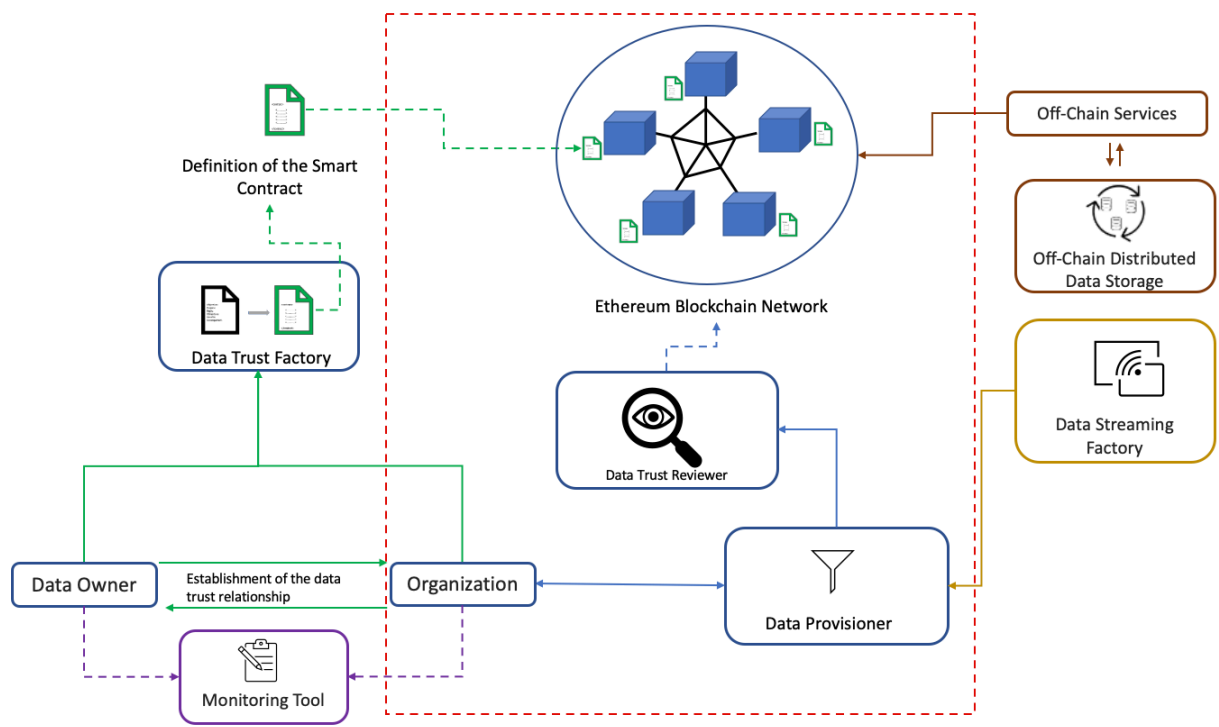

**Figure 4. Overview of the Architecture of the Data Trust Platform that Integrates Blockchain.**

One of the downsides of using Ethereum smart contracts to create a data trust platform is the cost of deploying smart contracts.
An alternative to save money would be managing the smart contracts in a centralized manner. A blockchain technology that enables this approach is Hyperledger. We can deploy in-chain code (equivalent to smart contracts) centrally. However, this approach affects the transparency that data trust demands.

## 6    Conclusions

This research proposes a hybrid data trust platform for IoT. IoT has unique characteristics that make some challenges emerge when implementing data trust – for instance, near real-time data streaming, low data rates, and heterogenous computing.

This research proposes the use of blockchain technology to enable data trust in IoT. The features of blockchain technology can guarantee the transparency, data value delivery, and consequence acceptance that data trust demands.

This research also presented a review of blockchain from different perspectives to help decide the most suitable blockchain configuration when implementing data trust in IoT. Public permission less blockchain technologies like Ethereum provide high transparency, value delivery, and consequence acceptance.

However, it provides low flexibility and might involve a high cost depending on the complexity of smart contracts.

On the other hand, Hyperledger Fabric chaincode might decrease costs and enable flexibility. However, as it is private and permissioned, it might guarantee low to medium transparency, value delivery, and consequence acceptance.

Implementation details and experiments of the proposed data trust architecture will be presented on a further research.